\documentclass[prd,aps,showpacs,nofootinbib,onecolumn,groupedaddress,amssymb,superscriptaddress]{revtex4}
\usepackage{amsmath}
\usepackage{amssymb}
\usepackage{amsfonts}
\usepackage{graphicx,bm}
\usepackage{dcolumn}
\usepackage[colorlinks=true]{hyperref}
\usepackage{color,amsxtra}
\usepackage{epsf}
\usepackage{enumerate}
\usepackage{hhline}
\usepackage{array}
\usepackage{tabularx}
\usepackage{subfigure}
\usepackage{fancyhdr}
\usepackage{mathrsfs}

\newcommand{\be}{\begin{equation}}
\newcommand{\ee}{\end{equation}}
\newcommand{\bea}{\begin{eqnarray}}
\newcommand{\eea}{\end{eqnarray}}
\newcommand{\beaa}{\begin{eqnarray*}}
\newcommand{\eeaa}{\end{eqnarray*}}

\newcommand{\nn}{\nonumber \\}

\def\be{\begin{equation}}
\def\ee{\end{equation}}
\def\bea{\begin{eqnarray}}
\def\eea{\end{eqnarray}}

\begin{document}
\title{Early Dark Energy with Power-law $F(R)$ Gravity}

\author{Sergei D. Odintsov}
\email{odintsov@ice.csic.es} \affiliation{Institut de Ci\`{e}ncies de l'Espai,
ICE/CSIC-IEEC, Campus UAB, Carrer de Can Magrans s/n, 08193 Bellaterra (Barcelona),
Spain}
 \affiliation{Instituci\'o Catalana de Recerca i Estudis Avan\c{c}ats (ICREA),
Passeig Luis Companys, 23, 08010 Barcelona, Spain}
\author{V.K. Oikonomou}
\email{voikonomou@gapps.auth.gr;v.k.oikonomou1979@gmail.com}
\affiliation{Department of Physics, Aristotle University of
Thessaloniki, Thessaloniki 54124, Greece}
\author{German~S.~Sharov}
 \email{sharov.gs@tversu.ru}
 \affiliation{Tver state university, Sadovyj per. 35, 170002 Tver, Russia}
 \affiliation{International Laboratory for Theoretical Cosmology,
Tomsk State University of Control Systems and Radioelectronics (TUSUR), 634050 Tomsk,
Russia}

\begin{abstract}
We study a power-law $F(R)$ gravity with an early dark energy
term, that can describe both the early-time and the late-time
acceleration of the Universe. We confront this scenario with
recent observational data including the Pantheon Type Ia
supernovae, measurements of the Hubble parameter $H(z)$ (Cosmic
Chronometers), data from Baryon Acoustic Oscillations and standard
rulers data from the Cosmic Microwave Background (CMB) radiation.
The model demonstrates some achievements in confronting with these
observations and can be compared with the
$\Lambda$-Cold-Dark-Matter model. In particular, in both models we
obtain  very close estimates for the Hubble constant $H_0$, but it
is not true for $\Omega_m^0$. The early dark energy term supports
viability of the considered $F(R)$ gravity model.
\end{abstract}


\maketitle

\section{Introduction}\label{Intr}

We live in an era in which the way to precision cosmology is paved
in rapid steps, aimed by the plethora of cosmological and
astrophysical observational data. There are a lot of facts and
challenges of standard benchmark models used currently in order to
describe the observable Universe. To be precise, the
$\Lambda$-Cold-Dark-Matter ($\Lambda$CDM) assisted with a nearly
scale invariant power spectrum observed in the Cosmic Microwave
Background (CMB), see to provide a generally acceptable
description of the current Universe at large scales. But there are
shortcomings, for example we do not even know if inflation indeed
was the source of the observed nearly scale invariant power
spectrum, if the inflationary era is controlled by a scalar field,
which has to be coupled to all the Standard Model particles in
order to reheat these, issues related to the inflationary era
solely. Regarding the $\Lambda$CDM, the possible existence of
large scale matter structure at large redshifts, will put further
into question the $\Lambda$CDM, which is a rather simple model,
not the actual model that describes the late-time Universe. To our
opinion, the $\Lambda$CDM is just a starting point, not the end.
There are many new issues to be addressed and much more new
physics to be employed from the primordial era, followed by the
reheating, matter, the CMB era and the late-time acceleration
eras. We have some hints about these eras, but we are still at the
start and in the next decades these model we have at hands, we
will be into test and the observations will reveal to us the
correct way to proceed in modelling the Universe. Modified
gravity, for example $f(R)$ gravity, is an alternative to dark
energy which can mimic the $\Lambda$CDM model and produce viable
late-time acceleration, in a geometric way. In some cases it can
also mimic dark matter, although there is strong motivation to
expect that dark matter is of particle nature, if not a massive
interacting particle, then some elusive light scalar field, like
the axion. Also modified gravity can model inflation in a
geometric way without the need for a scalar field to drive the
dynamics. The question is whether inflation ever existed at all.
In the next 15 years it will be scrutinized by stage 4 CMB
experiments and by gravitational waves experiments, so we will
have a concrete answer on whether it occurred or not. It should be
noted that the Universe's acceleration follows a
Friedmann-Robertson-Walker cosmology, but the latter remains after
all, an idealized solution of general relativity, which disregards
other interactions. Dark energy is not a straight consequence of
measuring results, but it follows from one possible
interpretation, among others. The real result of the measurements
is the detection of non-linear dependence of the registered energy
current density of SNe Ia with respect to redshift. Also the dark
Universe is currently not explained by the Standard-Model of
elementary particles which constitutes an embarrassing 96$\%$ of
the Universe. Thus the challenges for theorists are apparently
many, but this is exactly the core of the physics science, from
the date it was firstly quantified by Newton. Physics is always
challenged and we believe this is the correct way, physics always
in crisis, this is what keeps theorists active and busy. Each
generation will have another challenge, and the great wall of
understanding the Universe is built by each generation by putting
a simple brick in the wall. So expressions like physics in crisis
are redundant, physics was always in crisis and will be. For some
recent mainstream articles in the above research lines, see for
example
\cite{Spallicci:2020diu,Spallicci:2022jul,Sarracino:2022kve,Csaki:2001yk,Csaki:2001jk,Colin:2019opb,Dam:2017xqs}.
The Hubble tension problem is not a new problem, but it is a 30
years old problem, now very well constrained and verified at
5$\sigma$ C.L. and there exist various ways to remedy or alleviate
this problem
\cite{Capozziello:2020nyq,Spallicci:2021kye,Capozziello:2023ewq,Lopez-Corredoira:2022qdo}.
The tension between local low redshift \cite{HST2022} and CMB
measurements \cite{Planck:2018vyg} of the Hubble rate is nowadays
confirmed at $5\mathrm{\sigma}$, thus it is a realistic problem of
post recombination physics. To date it is still a mystery,
although many phenomenological solutions have been proposed that
can remedy or even alleviate the tension
\cite{Dai:2020rfo,He:2020zns,Nakai:2020oit,DiValentino:2020naf,Agrawal:2019dlm,Yang:2018euj,Ye:2020btb,Vagnozzi:2021tjv,
Desmond:2019ygn,OColgain:2018czj,Vagnozzi:2019ezj,
Krishnan:2020obg,Colgain:2019joh,Vagnozzi:2021gjh,Lee:2022cyh,Nojiri:2021dze,Krishnan:2021dyb,Ye:2021iwa,Ye:2022afu},
including the early dark energy (EDE) perspective
\cite{Niedermann:2020dwg,Poulin:2018cxd,Karwal:2016vyq,Oikonomou:2020qah,Nojiri:2019fft,KarwalEDE,EDE1,EDE2},
see also \cite{Kamionkowski:2022pkx} for a recent review and
several proposals for abrupt physics changes before $70-150\,$Myrs
\cite{Perivolaropoulos:2021jda,Perivolaropoulos:2021bds,Perivolaropoulos:2022vql,Odintsov:2022eqm,Odintsov:2022umu,Oikonomou:2022tjm}.
The $\Lambda$CDM model although being a benchmark model fitting
very well the CMB polarization anisotropies, it has its
shortcomings. However, the $\Lambda$CDM model is not the only one
which can describe in a consistent way the late-time acceleration
era. Modified gravity in its various forms
\cite{reviews1,reviews2,reviews3,reviews4} can also consistently
describe the late-time era, mimicking the $\Lambda$CDM model and
offering a solid theoretical background for model building. In
addition, modified gravity offers a theoretical framework which a
geometric fluid actually realizes dark energy but the same fluid
can also generate an inflationary era and also intermediate eras,
like an early dark energy era. Scenarios with an EDE are also
motivated by the $H_0$ tension problem. This form of dark energy
can be produced from scalar fields, axions or other forms of
matter, see for example a recent work on this
\cite{Oikonomou:2023bah}, but also from modified gravity. The EDE
component can play a role of dark energy or an effective
cosmological constant at intermediate times between the
matter-radiation equality  $z \simeq 3000$  and recombination $z
\simeq 1000$  and then decays faster than radiation. In a previous
work \cite{OdintsovSGS:2021} we considered an EDE generating
$F(R)$ gravity term as a possible tool for solving or alleviating
the $H_0$ tension problem. This model and the EDE term did not
affect significantly the values of $H_0$. In the present paper, we
explore a power-law $F(R)$ gravity model with a late-time
dominating term of the form $\sim\gamma R^\delta$ studied
previously in the papers \cite{Oikonomou:2020qah,Odintsov:2020nwm}
and with an additional EDE term firstly introduced in
Ref.~\cite{OdintsovSGS:2021}. We confront the model with the
following observational data: the Type Ia supernovae data (SNe Ia)
from the Pantheon sample survey, the Hubble parameter $H(z)$
measurements from differential ages of galaxies or cosmic
chronometers (CC), data connected with cosmic microwave background
radiation (CMB) and baryon acoustic oscillations (BAO). We shall
adopt the approach developed in some previous papers
\cite{OdintsovSGS:2020,Sharov:2016,PanSh:2017} and we obtain the
best fit parameters which in the end we shall compare with
predictions of the $\Lambda$CDM model. This paper is organized as
follows: In section \ref{Model}, we introduce the $F(R)$ model
under consideration, further, we consider its early-time evolution
in section~\ref{Inflation} and the late-time dynamics of the model
are analyzed in section~\ref{Late}. Section \ref{Observ} is
devoted to SNe Ia, $H(z)$, CMB and BAO observational data.  The
results of our analysis are presented in section \ref{Results} and
the conclusions follow at the end of the article.


\section{An $F(R)$ Gravity Model Quantifying Inflation, DE and EDE Eras}
 \label{Model}

We shall extend the $F(R)$ gravity model of Ref.
\cite{OdintsovSGS:2021}, with the general gravitational action
being,
 \begin{equation}
  S = \frac1{2\kappa^2}\int d^4x \sqrt{-g}\,F(R)  + S^{\mbox{\scriptsize matter}}\ ,
 \label{Act1}\end{equation}
where $\kappa^2=8\pi G$, $R$ is the Ricci scalar and
$S^{\mbox{\scriptsize matter}}$ is the matter action. In the
present paper, we analyze the power-law  $F(R)$ model of Ref.
\cite{OdintsovSGS:2021}  with an additional EDE term
\cite{OdintsovSGS:2021} so the $F(R)$ gravity function has the
form,
 \begin{equation}
 F(R)=R -2\Lambda\gamma \bigg(\frac{R}{2\Lambda}\bigg)^\delta +
  F_\mathrm{EDE}  + \frac{R^2}{M^2}\ .
  \label{FRde}
\end{equation}
The last term  $F_\mathrm{inf}=R^2/M^2 $ is responsible for the
early-time acceleration, \cite{Oikonomou:2020qah,Odintsov:2020nwm}
and dominates during the inflationary era, described below in
Sect.~\ref{Inflation}. However, the constant $M$ should be large
enough in order for $F_\mathrm{inf}$ to become negligible at late
times $z<3000$ related with our observational data.

The EDE term, $F_\mathrm{EDE}$, that mimics an effective
cosmological constant at intermediate times between the
matter-radiation equality and recombination was considered as
\cite{OdintsovSGS:2021}
 $$ 
    F_\mathrm{EDE}=-\alpha\cdot2\Lambda R_0^\ell
  \frac{R^{m-n}(R-R_0)^n}{R_0^{\ell+m}+R^{\ell+m}}\;,
 $$ 
 where $\alpha$, $\ell$, $m$, $n$, $R_0$ are  constants. The scale $R_0$ corresponds to the Ricci scalar value
for the  intermediate epoch $1000\le z\le 3000$. This EDE term
$F_\mathrm{EDE}$ was examined in Ref.~\cite{OdintsovSGS:2021} and
as it was shown it leads to a regular evolution without
singularities only if the numbers $\ell$, $m$, $n$ are $\ell=n=0$,
$m=1$ (if $\alpha$ is not negligible). So the viable form of the
EDE term is,
 \begin{equation}
 F_{\mathrm{EDE}}=-2\Lambda\alpha
 \frac{R}{R_0+R}\ . \label{FEDE}
\end{equation}
The EDE term (\ref{FEDE}) is aimed to shift the effective
cosmological constant before and near the time of recombination,
which might affect the CMB parameter measurements at that epoch.
At early times, where  $R\gg R_0$, the EDE term turns to be a
constant $F_{\mathrm{EDE}}\simeq -2\alpha\Lambda$ and becomes
irrelevant at the very early Universe. At late times $R\ll R_0$,
the EDE term becomes small because of the factor $R/R_0$.

Moreover, the EDE term (\ref{FEDE}) with sufficiently large
$\alpha$ can effectively suppress oscillations arising in this
model during the mentioned intermediate epoch and beyond
\cite{OdintsovSGS:2021}.

The field equations for $F(R)$ gravity with the action
(\ref{Act1}) are obtained by varying it with respect to the metric
$g_{\mu\nu}$:
$$ F_R R_{\mu\nu}-\frac F2
g_{\mu\nu}+\big(g_{\mu\nu}g^{\alpha\beta}\nabla_\alpha\nabla_\beta-\nabla_\mu\nabla_\nu\big)F_R
=\kappa^2T_{\mu\nu}\ .
 $$
Here $R_{\mu\nu}$ and $T_{\mu\nu}$ are the Ricci and
energy-momentum tensors respectively. In a spatially-flat
Friedman-Lema\^itre-Robertson-Walker (FLRW) spacetime with line
element,
 $$ds^2 =-dt^2 + a^2(t)\,d\mathbf{x}^ 2$$
where the scale factor is denoted as $a(t)$, these equations are
reduced to the system for the Hubble parameter $H=\dot a/a$, while
the Ricci scalar $R$ and the matter density $\rho$ are,
 \begin{eqnarray}
 \frac{dH}{d\log a}&=&\frac{R}{6H}-2H,\label{eqH}\\
 \frac{dR}{d\log a}&=&\frac1{F_{RR}}\bigg(\frac{\kappa^2\rho}{3H^2}-F_R+\frac{RF_R-F}{6H^2}\bigg),
 \label{eqR}\\
 \frac{d\rho}{d\log a}&=&-3(\rho+p)\, .  \label{cont}
 \end{eqnarray}
The matter density $\rho$ includes contributions of dust matter
$\rho_m$ (baryonic and dark matter) and radiation $\rho_r$. For
$\rho=\rho_m+\rho_r$   the continuity equation (\ref{cont}) can be
easily solved,
  \begin{equation}
 \rho=\rho_m^0a^{-3}+ \rho_r^0a^{-4}=\rho_m^0(a^{-3}+X_r a^{-4})\ .
 \label{rho}\end{equation}
Here $\rho_m^0$, $\rho_r^0$ and $a=1$ are the present time values
of the matter densities and the scale factor, while we assume the
estimation for the ratio of matter densities as provided by Planck
\cite{Planck:2018vyg}:
  \begin{equation}
  X_r=\frac{\rho_r^0}{\rho_m^0}=2.9656\cdot10^{-4}\ . \label{Xrm}
  \end{equation}

\section{Inflationary Evolution of the $F(R)$ Gravity Model}
 \label{Inflation}

In this section we briefly review the early time evolution of the
scenario (\ref{FRde}) that may be interpreted as the inflationary
era. For $F(R)$ models of the type (\ref{FRde}), in other words,
for models with the power-law ($\sim R^\delta$) and the
inflationary term  $F_\mathrm{inf}=R^2/M^2$ slow-roll inflation
scenarios were considered in many papers \cite{Odintsov:2020thl}.
These scenarios reproduce the inflationary era when the Ricci
scalar $R$ is very large, hence in the expression $F(R)$
(\ref{FRde}) we can neglect the terms,
  \begin{equation}
  F_{\mathrm{DE}}=-2\Lambda\gamma \bigg(\frac{R}{2\Lambda}\bigg)^\delta,\qquad
 F_{\mathrm{EDE}}=-2\Lambda\alpha
 \frac{R}{R_0+R}\simeq -2\alpha\Lambda,\quad R\gg R_0\,.
  \label{FDEEDE}
\end{equation}
The constant $M$ in $F_\mathrm{inf}$ may be evaluated as
 $ 
 M= 1.5 \cdot 10^{-5}\big(50/N\big)\,M_P
 $, 
where $N$ is the number of $e$-foldings  during inflation, and
$M_P =\sqrt{\frac{\hbar c}{4\pi G}} = 2.435\cdot10^{18}$ GeV is
the reduced Planck mass. Hence $M$ should be of order
$10^{13}\,$GeV that is close to approximate values of the Hubble
parameter $H(z)\simeq H_I \sim 10^{13}$ GeV during the
inflationary era in the mentioned scenarios
\cite{Oikonomou:2020qah,Odintsov:2020nwm}.

The considered inflationary scenario, is a slow-roll scenario,
assuming the approximation,
  \begin{equation}
|\ddot H|\ll H|\dot H|,\qquad |\dot H|\ll H^2.
 \label{slow}\end{equation}
The relation,
  \begin{equation}
R=6\dot H +12 H^2\, ,
 \label{RH}\end{equation}
equivalent to Eq.~(\ref{eqH}), in the slow-roll case (\ref{slow})
is reduced to $R\simeq 12 H^2$, hence the curvature scalar during
inflation is approximately $R\simeq R_I= {\cal O} (10^{27})$
GeV${}^2$. Consequently, we can evaluate the quadratic term
$F_\mathrm{inf}=R^2/M^2\simeq R_I^2/M^2= {\cal O} (10^{28})$
GeV${}^2$ during inflation (it is of order or exceeds $R\simeq
R_I$) and compare it with the terms (\ref{FDEEDE}). We estimate
the constant
  \begin{equation}
 \Lambda=3\Omega_\Lambda H_0^2\simeq4.2\cdot10^{-66}\mbox{ eV}^2
  \label{Lambda}\end{equation}
using the Planck 2018 fits \cite{Planck:2018vyg} for
$\Omega_\Lambda$ and $H_0$.  The term $F_{\mathrm{DE}}$ during
this era (for $R\simeq R_I$) is of order
$|F_{\mathrm{DE}}|\simeq8.4\cdot10^{-84}\gamma\cdot10^{110\delta}$
GeV${}^2$ and under natural limitations (see Sect.~\ref{Results})
$\gamma\le10$, $\delta\le0.06$ it does not exceed
$|F_{\mathrm{DE}}|\simeq3.4\cdot10^{-76}$ GeV${}^2$.

Taking into account the EDE term
$|F_{\mathrm{EDE}}|\simeq8.4\cdot10^{-84}\alpha$ GeV${}^2$, we
obtain the relations of the terms (\ref{FDEEDE}) to
$F_\mathrm{inf}$ near the inflationary epoch,
  \begin{equation}
  \frac{|F_{\mathrm{DE}}|}{F_\mathrm{inf}}<5\cdot10^{-104},\qquad
 \frac{|F_{\mathrm{EDE}}|}{F_\mathrm{inf}}<10^{-106}\,,
  \label{FDEinf}
\end{equation}
 if $\gamma\le10$, $\delta\le0.06$, $\alpha\le10^5$. Derivatives of
$F_{\mathrm{DE}}$ and $F_{\mathrm{EDE}}$ are relatively smaller.

  We may conclude that the terms (\ref{FDEEDE}) are negligible during
the inflationary era, when $F(R)$ is reduced to the form,
  \begin{equation}
 F(R)\simeq R + \frac{R^2}{M^2}\;.
  \label{FRInf}
\end{equation}
If we use the relation (\ref{RH}), we can exclude $R$ and rewrite the equation
(\ref{eqR}) for the case (\ref{FRInf}) neglecting matter:
  \begin{equation}
\ddot H+3H\dot H-\frac{\dot H^2}{2H}+\frac{M^2}{12}H=0\, .
 \label{eqH2}\end{equation}
Under the slow-roll conditions (\ref{slow}) this equation is simplified to
 $\dot H\simeq-M^2/36$, that yields a quasi-de Sitter evolution
  \begin{equation}
H(t) = H_I - \frac{M^2}{36} t
 \label{qdS}\end{equation}
and the slow-roll parameter $\epsilon_1$ reads
 \cite{Odintsov:2020thl},
  \begin{equation}
\epsilon_1 = -\frac{\dot H}{H^2}= \frac{M^2}{36\big(H_I -
\frac{M^2}{36} t\big)^2}\, .
 \label{ep1}\end{equation}
We can find the time $t_f$, when inflation ends, from the equality
$\epsilon_1(t_f) = 1$, that yields $t_f=6(6H_I- M)/M^2$. The
horizon crossing time instance $t_i$ for inflation  we can express
via the $e$-folding number $N$ \cite{Odintsov:2020thl}
  \begin{equation}
N = \int_{t_i}^{t_f}H(t)\,dt=H_I(t_f-t_i)-\frac{M^2}{72}(t_f-t_i)^2
 \label{Nfold}\end{equation}
and obtain $t_i=\frac{36}{M^2}H_I- \frac 6M\sqrt{2N+1}$. If we
evaluate the slow-roll index $\epsilon_1$ (\ref{ep1}) at the time
instance $t_i$, we get the result \cite{Odintsov:2020thl}
  \begin{equation}
\epsilon_1(t_i) = \frac1{2N+1}\,.
 \label{ep1N}\end{equation}
The other relevant slow-roll parameters have the form
\cite{Odintsov:2020thl},
 $$
\epsilon_3 = \frac{\dot F_R}{2HF_R},\qquad \epsilon_4 = \frac{\dot F_R}{2HF_R},
 $$
can also be calculated for the evolution (\ref{qdS}) at $t=t_i$:
   \begin{equation}
\epsilon_3(t_i) = -\frac1{2N+2},\qquad \epsilon_4(t_i) = -\frac1{2N+1}
 \label{ep23}\end{equation}
in concordance with the estimations \cite{Odintsov:2020thl}
 $\epsilon_4\simeq-\epsilon_1(1+24H^2\frac{F_{RRR}}{F_{RR}})$,
 $\epsilon_3=-\epsilon_1/(1-\epsilon_4)$. Here for all $\epsilon_i$ we have
 $|\epsilon_i|\ll1$. If we calculate the spectral index of primordial curvature perturbations $n_s$ and the
tensor-to-scalar ratio $r$ \cite{Odintsov:2020thl}
 $$ 
n_s=1-\frac{4\epsilon_1-2\epsilon_3+2\epsilon_4}{1-\epsilon_1},\qquad r=48
\frac{\epsilon_3^2}{(1+\epsilon_3)^2},
 $$ 
for the expressions (\ref{ep1N}), (\ref{ep23}),  we get the estimates
   \begin{equation}
n_s=1-\frac 2N+ \frac1{2N(N+1)},\qquad r=\frac{48}{(2N+1)^2}.
 \label{nsr}\end{equation}
They are close to the well-known results
$n_s\simeq1-4\epsilon_1\simeq1-2/N$, $r\simeq12/N^2$ for the
spectral index and the tensor-to-scalar ratio. One may conclude
that the inflationary behavior of the model (\ref{FRde}) is viable
and compatible with the Planck 2018 data \cite{Planck:2018vyg}.

\section{Late-time Evolution of the $F(R)$ Gravity Model}
 \label{Late}

During the late-time evolution, in other words, at redshifts $z\le
3000$, the EDE term $F_{\mathrm{EDE}}$ (\ref{FEDE}) and matter
play an important role, and we shall see below that the
inflationary term $F_\mathrm{inf}$ becomes negligible.

It was shown in the paper  \cite{OdintsovSGS:2021} that the model
(\ref{FRde}), (\ref{FEDE}) under certain initial conditions mimics
the $\Lambda$CDM model at large redshifts $z>1000$ with the
following asymptotic behavior of $H$ and $R$:
\begin{equation}
 \frac{H^2}{(H^{*}_0)^2}=\Omega_m^{*} \big(a^{-3}+ X_ra^{-4}\big)+\Omega_\Lambda^{*},\qquad
 \frac{R}{2\Lambda}=2+\frac{\Omega_m^{*}}{2\Omega_\Lambda^{*}}a^{-3}, \qquad
 a\to0.
  \label{asymLCDM}\end{equation}
Here the matter density $\rho=\rho_m+\rho_r$ behaves as (\ref{rho}), (\ref{Xrm}); the
index $*$ refers to parameters related to the $\Lambda$CDM model, in particular,
$H^{*}_0$ is the Hubble constant in the $\Lambda$CDM scenario. However, under these
initial conditions the late-time evolution for the $F(R)$ model deviates from the
$\Lambda$CDM evolution, such that the parameters $H_0=H(t_0)$,
$\Omega_m^0=\frac{\kappa^2}{3H_0^2}\rho_m(t_0)$ measured today (at $t_0$) for our models
will be different:
 $$
 H_0\ne H^{*}_0, \qquad \Omega_m^0\ne \Omega_m^{*}\ .
 $$
However, these parameters are connected:
 \begin{equation}
 \Omega_m^0H_0^2=\Omega_m^{*}(H^{*}_0)^2=\frac{\kappa^2}3\rho_m(t_0),
 \qquad  \Omega_\Lambda H_0^2=\Omega_\Lambda^{*}(H^{*}_0)^2=\frac{\Lambda}3\ .
  \label{H0Omm}\end{equation}
Using the relations (\ref{asymLCDM}) and (\ref{Lambda}) we can estimate the Ricci scalar
and the inflationary term for $z\sim 3000$:
 $$\frac{R}{2\Lambda}\sim3\cdot10^{10},\qquad R\sim 10^{-73}\mbox{ GeV}^2,\qquad
  F_\mathrm{inf}=\frac{R^2}{M^2} \sim 10^{-172}\mbox{ GeV}^2.
 $$
One can see that $F_\mathrm{inf}$ is many orders smaller than $R$
and  the terms (\ref{FDEEDE})
$|F_\mathrm{DE}|>\gamma\cdot10^{-83}\mbox{ GeV}^2$,
$|F_\mathrm{EDE}|\sim\alpha\cdot10^{-83}\mbox{ GeV}^2$. Later,
during further evolution, $F_\mathrm{inf}$ diminishes faster than
other terms.

Following Refs.~\cite{OdintsovSGS:2021} we redefine the Hubble
parameter and the Ricci scalar as dimensionless functions,
\begin{equation}
E=\frac{H}{H_0^{*}},\qquad  {\cal R}=\frac{R}{2\Lambda}\ .
   \label{ER}\end{equation}
If we neglect the inflationary term $F_\mathrm{inf}$, the
dynamical equations (\ref{eqR}) for the model (\ref{FRde}),
(\ref{FEDE}) with the dimensionless variables (\ref{ER}) can be
rewritten as:
\begin{eqnarray}
\frac{dE}{d\log a}&=&\Omega_\Lambda^{*}\frac{{\cal R}}{E}-2E\ , \nn   
\frac{d{\cal R}}{d\log a}&=&\frac1{RF_{RR}}
 \bigg[\frac{\Omega_m^{*}(a^{-3}+ X_ra^{-4})+\Omega_\Lambda^{*}\big[\gamma(1-\delta){\cal R}^{\delta}+
 \alpha\frac{{\cal R}^2}{({\cal R}_0+{\cal R})^2}\big]}{E^2}-
 1+\gamma\delta{\cal R}^{\delta-1}+ \alpha\frac{{\cal R}_0}{({\cal R}_0+{\cal R})^2}\bigg]\ .
  \label{eqRde}\end{eqnarray}
Here $RF_{RR}=\gamma\delta(1-\delta){\cal
R}^{\delta-1}+2\alpha{\cal R}_0{\cal R}/({\cal R}_0+{\cal R})^3$,
 ${\cal R}_0=R_0/(2\Lambda)$. As mentioned above, the model  (\ref{FRde}), (\ref{FEDE}) mimics
the $\Lambda$CDM asymptotic behavior (\ref{asymLCDM}) at large
redshifts or at the range $10\le{\cal R}\le10^{10}$ of the
normalized Ricci scalar \cite{OdintsovSGS:2021}. We can
numerically solve the system of equations (\ref{eqRde}) by
integrating over the folding number $x=\log a=-\log(z+1)$ and
assuming the initial conditions (\ref{asymLCDM}) at an initial
point $x_i$ related to redshift $z_i$ in the range $1000\le z \le
3000$. This redshift range should also contain $z_0$ corresponding
to the value $R_0$ of the EDE term (\ref{FEDE}): $R(z_0)=R_0$. At
this time, just before recombination and at its vicinity (where
$R\sim R_0$), the EDE term becomes important. In our calculations
we chose ${\cal R}_0=10^8$ and the initial point $x_i$ before
$x_0=-\log(z_0+1)$. In this case the results appeared to be weakly
depending on $x_i$.

In the paper \cite{OdintsovSGS:2021} we studied solutions of the
system (\ref{eqRde}) without the EDE term ($\alpha=0$) and had
seen their undesirable oscillatory behavior at large $R$ (see also
Refs.~\cite{Oikonomou:2020qah,Odintsov:2020nwm}). These
oscillations become inevitable in the most interesting case
$\delta\ll 1$. However, the EDE term (\ref{FEDE}) with
sufficiently large $\alpha$ makes it possible to suppress these
oscillations in the framework of the considered model.

Thus, we can obtain non-oscillating and non-diverging solutions of the model
(\ref{FRde}), (\ref{FEDE}), and, further, confront them with the observational data
(Sect.~\ref{Observ}) by fitting the free parameters. If we fix the ratio $X_r$
(\ref{Xrm}) and the values ${\cal R}_0$ and $x_i$, we will work with the following set
of the parameters:
 \be
\gamma,\;\delta,\;\alpha,\;\Omega_m^*, \; \Omega_\Lambda^{*},\; H_0^*\,.
 \label{param}\ee
The last 3 parameters should be transformed into the more convenient $\Omega_m^0$,
$\Omega_\Lambda$, $H_0$ via the relations (\ref{H0Omm}) and $H_0=H_0^*E(z=0)$. Here we
keep in mind that the calculated normalized value $E(z)$ yields the Hubble parameter as
$H(z)=H_0^*E(z)$.

\section{Observational data}\label{Observ}

We confront the model (\ref{FRde}), (\ref{FEDE}) and its
predictions  with the observational data in order to obtain the
best fit for the free parameters (\ref{param}) and estimate
viability of the model.  In this paper our analysis involves the
following observational data: (a) Pantheon sample of Type Ia
supernovae (SNe Ia) data \cite{Scolnic17}; (b) measurements of the
Hubble parameter $H(z)$ from cosmic chronometers, data from (c)
cosmic microwave background radiation (CMB)  and (d) baryon
acoustic oscillations (BAO). Unlike Ref.~\cite{OdintsovSGS:2020}
here we include the
renewed BAO data and exclude $H(z)$ estimations extracted from BAO.\\

The Pantheon sample database \cite{Scolnic17} for SNe Ia contains
$N_{\mathrm{SN}}=1048$ datapoints of distance moduli
$\mu_i^\mathrm{obs}$ at redshifts $z_i$. We calculate the $\chi^2$
function,
 \begin{equation}
\chi^2_{\mathrm{SN}}(\theta_1,\dots)=\min\limits_{H_0} \sum_{i,j=1}^{N_\mathrm{SN}}
 \Delta\mu_i\big(C_{\mathrm{SN}}^{-1}\big)_{ij} \Delta\mu_j,\qquad
 \Delta\mu_i=\mu^\mathrm{th}(z_i,\theta_1,\dots)-\mu^\mathrm{obs}_i\ ,
 \label{chiSN}\end{equation}
where  $\theta_j$ are free model parameters, $C_{\mbox{\scriptsize
SN}}$ is the covariance matrix  \cite{Scolnic17} and the
theoretical values of distance moduli are given by the relations,
 \bea
 \mu^\mathrm{th}(z)
&=& 5 \log_{10} \frac{D_L(z)}{10\mbox{pc}}, \qquad D_L (z)= (1+z)\, D_M(z),\nonumber\\
D_M(z)&=& c \int\limits_0^z\frac{d\tilde z}{H(\tilde
 z)}. \label{DM}\eea
In the $\chi^2$ function (\ref{chiSN}) for SNe Ia data we consider
the Hubble constant $H_0$ or the ``asymptotical'' constant $H_0^*$
(\ref{asymLCDM}), (\ref{H0Omm}) as a nuisance parameter following
Refs.~\cite{OdintsovSGS:2020,OdintsovSGS:2021}. For the Hubble
parameter data  $H(z)$ here we use the cosmic chronometers (CC),
i.e. measured different ages $\Delta t$ of galaxies with close
redshifts $\Delta z$ and< consequently, extracted values $
 H (z)= \frac{\dot{a}}{a} \simeq -\frac{1}{1+z} \frac{\Delta z}{\Delta t}.
 $

In this paper, we also consider $N_H=32$ CC $H(z)$ data points, 31
of them are given in Refs.~\cite{HzData} and were used previously
in the papers \cite{OdintsovSGS:2020,OdintsovSGS:2021}. Here we
add the value $H=98.8\pm33.6$ at $z=0.75$ from
Ref.~\cite{Borghi:2022}. Unlike Ref.~\cite{OdintsovSGS:2021} we do
not include here $H(z)$ values, extracted from BAO data along the
line-of-sight direction to avoid correlations with BAO data
described below. We calculate the $\chi^2$ function for CC $H(z)$
data as follows:
\begin{equation}
    \chi_H^2(\theta_1,\dots)=\sum_{j=1}^{N_H}\bigg[\frac{H(z_j,\theta_1,\dots)-H^{obs}(z_j)}{\sigma _j}  \bigg]^2
    \label{chiH}
\end{equation}
We use observational manifestations from the CMB radiation that
are extracted from Planck 2018 data \cite{Planck:2018jri} as  the
following parameters \cite{ChenHuangW2018}:
  \begin{equation}
  \mathbf{x}=\big(R,\ell_A,\omega_b\big),\qquad R=\sqrt{\Omega_m^0}\frac{H_0D_M(z_*)}c,\quad
 \ell_A=\frac{\pi D_M(z_*)}{r_s(z_*)},\quad\omega_b=\Omega_b^0h^2\ .
 \label{CMB} \end{equation}
Here, $z_*$ is the photon-decoupling redshift, $D_M$ is the
comoving distance (\ref{DM}),
$h=H_0/[100\,\mbox{km}\mbox{s}^{-1}\mbox{Mpc}^{-1}]$, the comoving
sound horizon $r_s(z)$ is calculated as,
  \begin{equation}
r_s(z)=  \int_z^{\infty} \frac{c_s(\tilde z)}{H (\tilde z)}\,d\tilde
z=\frac1{\sqrt{3}}\int_0^{1/(1+z)}\frac{da}
 {a^2H(a)\sqrt{1+\big[3\Omega_b^0/(4\Omega_\gamma^0)\big]a}}\ .
  \label{rs2}\end{equation}
We estimate the ratio of baryons and photons
$\Omega_b^0/\Omega_\gamma$ using the relation (\ref{Xrm}),
$\rho_\nu=N_\mathrm{eff}(7/8)(4/11)^{4/3}\rho_\gamma$ with
$N_\mathrm{eff} = 3.046$, as given by Planck 2018 data
\cite{Planck:2018jri}. We use the estimation of $z_*$  given in
Refs.~\cite{ChenHuangW2018,HuSugiyama95}. The current baryon
fraction $\Omega_b^0$ here is considered as the nuisance parameter
in the corresponding $\chi^2$ function,
 \begin{equation}
\chi^2_{\mbox{\scriptsize CMB}}=\min_{\omega_b}\Delta\mathbf{x}\cdot
C_{\mathrm{CMB}}^{-1}\big(\Delta\mathbf{x}\big)^{T},\qquad \Delta
\mathbf{x}=\mathbf{x}-\mathbf{x}^{Pl}\ .
 \label{chiCMB} \end{equation}
The estimates~\cite{ChenHuangW2018},
  \begin{equation}
  \mathbf{x}^{Pl}=\big(R^{Pl},\ell_A^{Pl},\omega_b^{Pl}\big)=\big(1.7428\pm0.0053,\;301.406\pm0.090,\;0.02259\pm0.00017\big)\ ,
   \label{CMBpriors} \end{equation}
are extracted from Planck collaboration 2018 data
\cite{Planck:2018jri} with free amplitude for the lensing power
spectrum. The covariance matrix $C_{\mathrm{CMB}}=\|\tilde
C_{ij}\sigma_i\sigma_j\|$ is described in
Ref.~\cite{ChenHuangW2018}.

For the baryon acoustic oscillations (BAO) data we consider the
magnitudes,
\begin{equation}
d_z(z)= \frac{r_s(z_d)}{D_V(z)}\, ,\qquad A(z) = \frac{H_0\sqrt{\Omega_m^0}}{cz}D_V(z)\,
, \label{dzAz}
\end{equation}
where $D_V(z)=\big[{cz D_M^2(z)}/{H(z)} \big]^{1/3}$,  $r_s(z_d)$
is the integral (\ref{rs2}) and $z_d$ being the redshift at the
end of the baryon drag era. Here we work with 21 BAO data points
for $d_z(z)$ and 7 data points for $A(z)$ as given in
Table~\ref{TBAO} from
Refs.~\cite{Percival:2009,Kazin:2009,Beutler:2011,Blake:2011,Chuang:2013,Anderson:2013,Ross:2014,Beutler:2016,Chuang:2017,Bourboux:2017,
Zhu:2018,Blomqvist:2019,Hou:2020,Tamone:2020}. This table contains
some new data points with respect to BAO data from
 Refs.~\cite{OdintsovSGS:2020}. However we
excluded from Table~\ref{TBAO} estimates of $d_z$, extracted from
repeating or overlapping galaxy catalogues.
\begin{table}[th]
\centering
 {\begin{tabular}{||l|l|l|l|l|l|l||}
\hline
 $z$  & $d_z(z)$ &$\sigma_d$    & $ A(z)$ & $\sigma_A$   & Survey & Refs.\\ \hline
 0.106& 0.336  & 0.015 & 0.526& 0.028&  6dFGS & \cite{Beutler:2011}\\ \hline
 0.15 & 0.2237 & 0.0084& -    & -    &  SDSS DR7&\cite{Ross:2014} \\ \hline
 0.20 & 0.1905 & 0.0061& 0.488& 0.016&  SDSS DR7&\cite{Percival:2009} \\ \hline
 0.278& 0.1394 & 0.0049& -    & -    &  SDSS LRG&\cite{Kazin:2009} \\ \hline
 0.314& 0.1239 & 0.0033& -    & -    & SDSS LRG &\cite{Blake:2011} \\ \hline
 0.32 & 0.1181 & 0.0026& -    & -    &   DR10,11&\cite{Anderson:2013} \\ \hline
 0.32 & 0.1165 & 0.0024& -    & -    & BOSS DR12&\cite{Chuang:2017} \\ \hline
 0.35 & 0.1097 & 0.0036& 0.484& 0.016& SDSS DR7 &\cite{Percival:2009} \\ \hline
 0.38 & 0.1011 & 0.0011& -    & -    & BOSS DR12&\cite{Beutler:2016} \\ \hline
 0.44 & 0.0916 & 0.0071& 0.474& 0.034&  WiggleZ &\cite{Blake:2011}\\ \hline
 0.57 & 0.0739 & 0.0043& 0.436& 0.017&  BOSS DR9&\cite{Chuang:2013}\\ \hline
 0.57 & 0.0726 & 0.0014& -    & -    &  DR10,11 &\cite{Anderson:2013}\\ \hline
 0.59 & 0.0701 & 0.0008& -    & -    & BOSS DR12&\cite{Chuang:2017} \\ \hline
 0.60 & 0.0726 & 0.0034& 0.442& 0.020&  WiggleZ &\cite{Blake:2011}\\ \hline
 0.61 & 0.0696 & 0.0007& -    & -    & BOSS DR12&\cite{Beutler:2016} \\ \hline
 0.73 & 0.0592 & 0.0032& 0.424& 0.021& WiggleZ  &\cite{Blake:2011} \\ \hline
 0.85 & 0.0538 & 0.0041& -& - &  DR16 ELG       &\cite{Tamone:2020}\\ \hline
 1.48 & 0.0380 & 0.0013& -& - & eBOSS DR16      &\cite{Hou:2020}\\ \hline
 2.0  & 0.0339 & 0.0025& -& - & eBOSS DR14      &\cite{Zhu:2018}\\ \hline
 2.35 & 0.0327 & 0.0016& -& - & DR14 Ly$\alpha$ &\cite{Blomqvist:2019}\\ \hline
 2.4  & 0.0331 & 0.0016& -& - & DR12 Ly$\alpha$ &\cite{Bourboux:2017}\\  \hline
 \end{tabular}
 \caption{BAO data $d_z(z)=r_s(z_d)/D_V(z)$ and $A(z)$ (\ref{dzAz}).}
 \label{TBAO}} \end{table}
The  $\chi^2$ function yields:
\begin{equation}
\chi^2_{\mathrm{BAO}}(\Omega_m^0,\theta_1,\dots)=\Delta d\cdot C_d^{-1}(\Delta d)^T +
\Delta { A}\cdot C_A^{-1}(\Delta { A})^T\, . \label{chiBAO}
\end{equation}
Here, $C_{d}$ and $C_{A}$ are the covariance matrices for the correlated BAO data
\cite{Percival:2009,Blake:2011} and the corresponding vectors are:
 $$
\Delta d_i=d_z^\mathrm{obs}(z_i)-d_z^\mathrm{th}(z_i,\dots)\, , \quad \Delta
A_i=A^\mathrm{obs}(z_i)-A^\mathrm{th}(z_i,\dots)\, .
 $$

\section{Results and discussion}\label{Results}

\begin{figure}[bh]
  \centerline{\includegraphics[scale=0.71,trim=4mm 4mm 5mm 5mm]{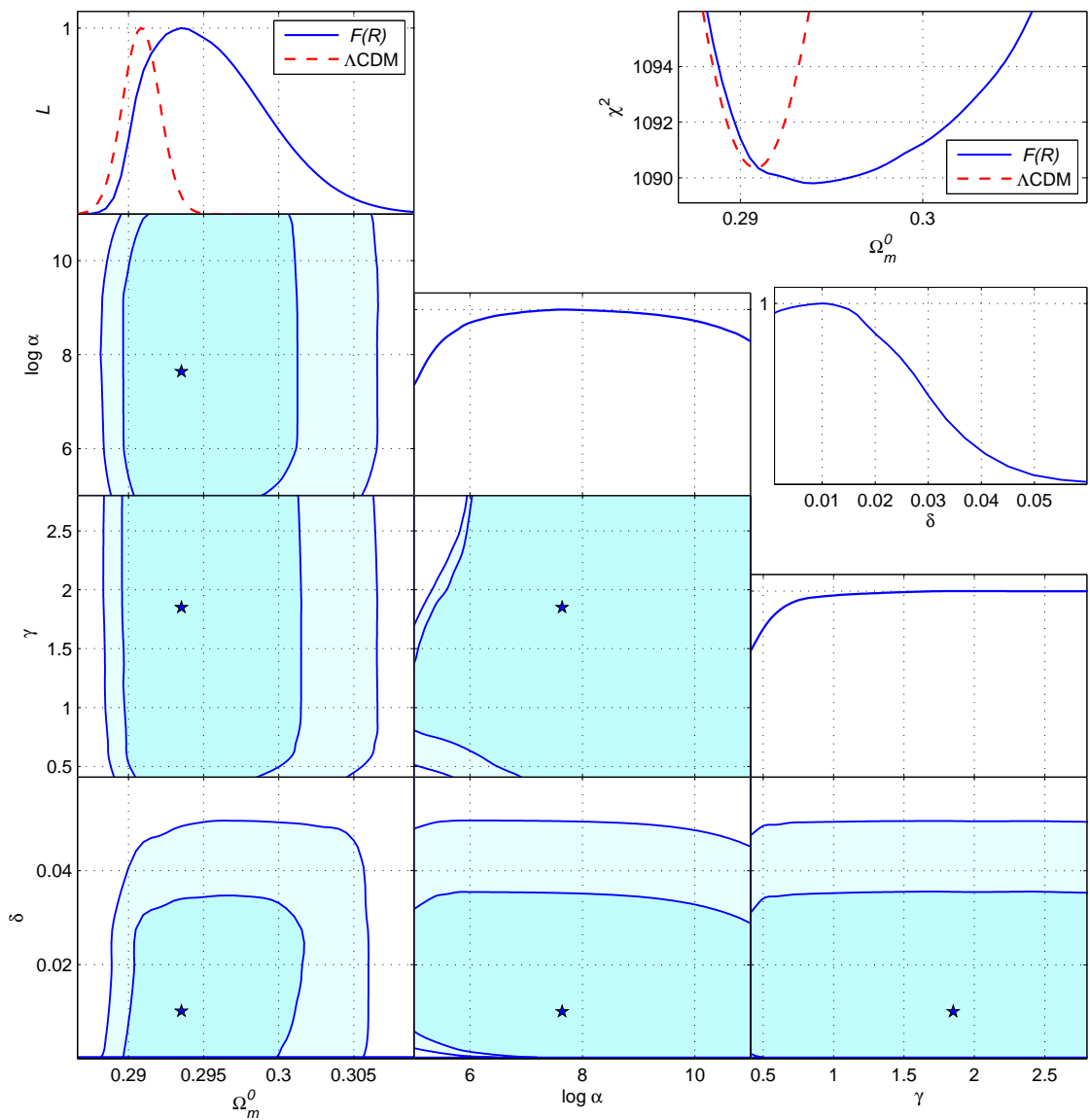}}
\caption{Contour plots of $\chi^2_\mathrm{tot}$ with $1\sigma$, $2\sigma$ CL and
likelihood functions $ {\cal L}(\theta_i)$ for the $F(R)$ model (\ref{FRde}),
(\ref{FEDE}) in comparison with  the $\Lambda$CDM model in one-parameter distributions
$\chi^2_\mathrm{tot}(\Omega_m^0)$ and  $ {\cal L}(\Omega_m^0)$. }
  \label{F1}
\end{figure}

Our $F(R)$ model (\ref{FRde}), (\ref{FEDE}) will fit the observational data from the
previous section, if we minimize the $\chi^2$ function, including SNe Ia
 (\ref{chiSN}), CC $H(z)$ data (\ref{chiH}), CMB (\ref{chiCMB}) and BAO (\ref{chiBAO})
 contributions:
   \begin{equation}
  \chi^2_\mathrm{tot}=\chi^2_\mathrm{SN}+\chi^2_H+\chi^2_\mathrm{CMB}+\chi^2_\mathrm{BAO}\,.
 \label{chitot} \end{equation}
This function is calculated in the space of free model parameters
(\ref{param}) with flat priors within their natural limitations
(positive values for all of them). The corresponding contour plots
and likelihoods for $\gamma$, $\delta$, $\alpha$, $\Omega_m^0$ are
shown in Fig.~\ref{F1}.
The  contour plots are depicted as $1\sigma$ (68.27\%) and $2\sigma$ (95.45\%)
confidence level domains for two-parameter distributions
$\chi^2_\mathrm{tot}(\theta_i,\theta_j)$, assuming minimization over all remaining free
model parameters with the mentioned natural limitations. For example, in the bottom-left
panel in Fig.~\ref{F1} in $\Omega_m^0-\delta$ plane we present $1\sigma$  and $2\sigma$
CL contours for
$\chi^2_\mathrm{tot}(\Omega_m^0,\delta)=\min\limits_{\gamma,\alpha,\Omega_\Lambda^{*},
H_0}\chi^2_\mathrm{tot}$. In other panels in Fig.~\ref{F1} we use the similar approach
with $\chi^2_\mathrm{tot}(\theta_i,\theta_j)$, these functions reach their absolute
minima at the points marked by stars.

Note that we use the ``true'' parameter
$\Omega_m^0=\frac{\kappa^2}{3H_0^2}\rho_m(t_0)$ in the left panels
of  Fig.~\ref{F1} and in the top-right panel, where the
one-parameter distribution $\chi^2_\mathrm{tot}(\Omega_m^0)$ is
shown, in comparison with its analog for the $\Lambda$CDM model.
Here one-parameter distributions are also minimized over all
remaining model parameters. The value $\Omega_m^0$ should be
differed from the ``asymptotic'' parameter  $\Omega_m^*$
(\ref{asymLCDM}) keeping in mind their connection
 (\ref{H0Omm}).  For $\gamma-\delta$ or $\log\alpha-\gamma$ planes the values  $\Omega_m^*$,
 $\Omega_\Lambda^*$, $H_0^*$ are more convenient, when we minimize  $\chi^2_\mathrm{tot}$
over these parameters.

We see in Fig.~\ref{F1} that the best fitted value
$\Omega_m^0=0.294^{+0.0048}_{-0.0036}$ for the $F(R)$ model is larger and has more wide
$1\sigma$ error box than this value in the $\Lambda$CDM model. This is connected, in
particular, with large number $N_p=6$ of its model parameters (\ref{param}) in
comparison with $N_p=2$ in the $\Lambda$CDM model.

For $\Omega_m^0$, $\gamma$, $\delta$, $\alpha$  the likelihood
functions are depicted in Fig.~\ref{F1}. They are connected with
the corresponding one-parameter distributions, in particular,
 ${\cal L}(\Omega_m^0)\sim\exp(-\chi^2_\mathrm{tot}(\Omega_m^0)/2)$.

\begin{figure}[hb]
  \centerline{\includegraphics[scale=0.68,trim=6mm 4mm 5mm 1mm]{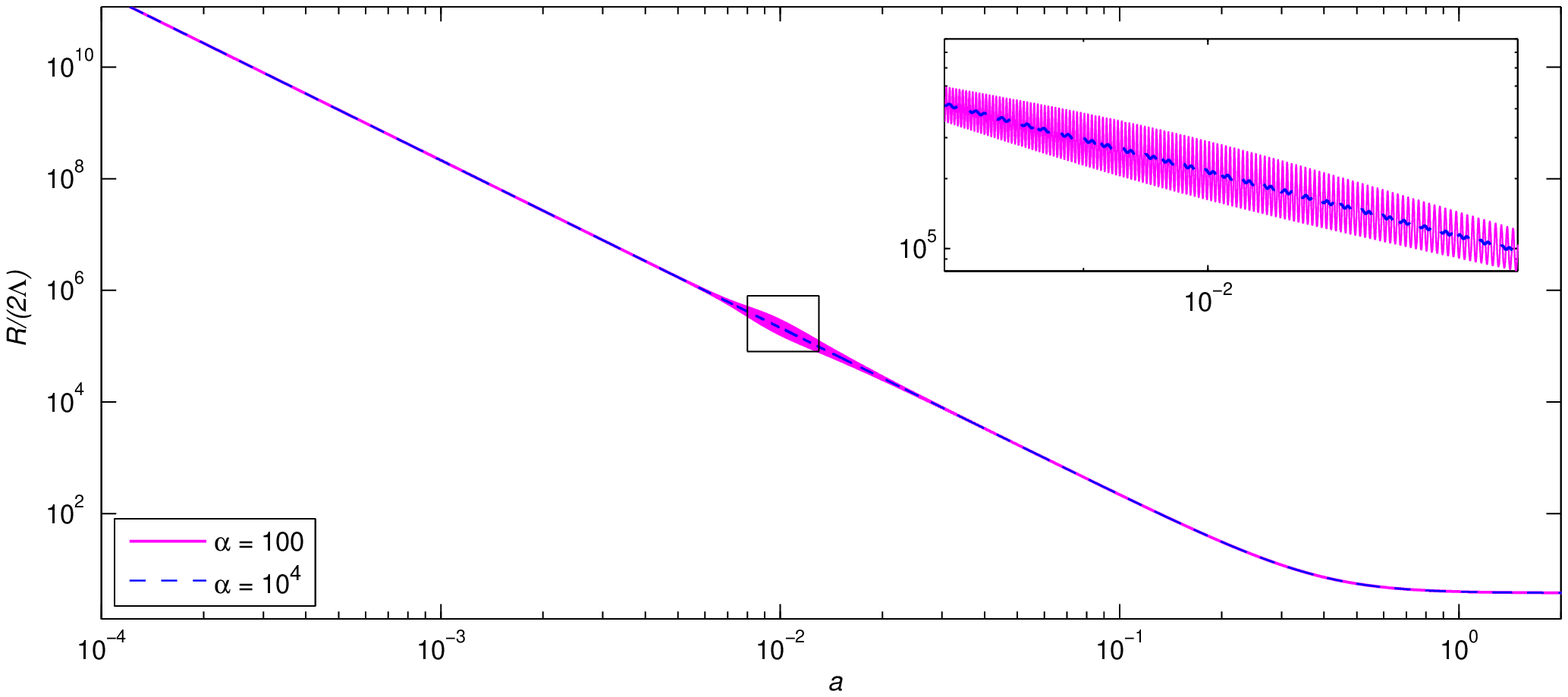}}
\caption{Oscillations in the model (\ref{FRde}) and their suppression at high $\alpha$.}
  \label{F2}
\end{figure}

The best fit $1\sigma$ estimates for other parameters and values
$\min\chi^2_\mathrm{tot}$ are tabulated in Table \ref{Estim}. They
are determined from the one-parameter distributions or likelihoods
${\cal L}(\theta_j)$. We see that the $F(R)$ model (\ref{FRde}),
(\ref{FEDE}) has the small advantage over the $\Lambda$CDM model
in $\min\chi^2_\mathrm{tot}$, but this advantage disappears if we
take into account numbers  $N_p=6$ and $N_p=2$ of free model
parameters and consider the Akaike information criterion
\cite{Akaike} $\mbox{AIC} = \min\chi^2_{tot} +2N_p$.
\begin{table}[ht]
\begin{tabular}{|l|c|c|c|c|c|c|c|}
\hline  Model &   $\min\chi^2_\mathrm{tot}/d.o.f$& AIC & $\Omega_m^0$& $H_0$ &  $\delta$ & $\gamma$  & $\log\alpha$  \\
\hline
 $F(R)$+EDE & 1089.80 /1106 & 1101.80& $0.294^{+0.0048}_{-0.0036}$ & $68.93^{+1.61}_{-1.57}$& $0.010^{+0.017}_{-0.010}$ & $1.85^{+6.3}_{-1.51}$ & $7.64^{+4.45}_{-2.62}$ \rule{0pt}{1.1em}  \\
\hline
$\Lambda$CDM& 1090.35 /1110 & 1094.35& $0.2908^{+0.0013}_{-0.0012}$& $68.98^{+1.58}_{-1.60}$& - &- &- \rule{0pt}{1.1em}  \\
\hline
 \end{tabular}
 \caption{The best  fit values for parameters and $\min\chi^2_\mathrm{tot}$ for the $F(R)$ model
(\ref{FRde}), (\ref{FEDE})  in comparison with the $\Lambda$CDM model.}
\label{Estim}
\end{table}
In Fig.~\ref{F1} we use the more convenient parameter $\log\alpha$
instead of the EDE factor $\alpha$, because the $1\sigma$ error
box includes rather large values
$\alpha\in[1.5\cdot10^2,1.78\cdot10^5]$, whereas the best fit is
$\alpha\simeq2.1\cdot10^3$. If the EDE factor $\alpha$ is lower
than a certain permissible value, oscillations appear during the
middle-time evolution in the considered model (\ref{FRde}),
(\ref{FEDE}). An example of these oscillations for the Ricci
scalar $R=R(a)$ is presented in Fig.~\ref{F2}. Here other model
parameters are close to  best fit values from Table~\ref{Estim}.
We see that for $\alpha=100$ the value $R(a)$ begins to oscillate
near the middle-time epoch $z\sim 100$ ($a\sim 10^{-2}$), that is
later than the recombination and the initial point $z_i$, where we
start integration of the system (\ref{eqRde}). For  $\alpha=10^4$
and the same other parameters  these oscillations appear to be
suppressed.

If we take values $\alpha$ lower than 100 for the considered in
Fig.~\ref{F2} system, or put $\alpha=0$ (that is exclude the EDE
term), we will obtain oscillations with extremely growing
amplitudes. In other words, smooth solutions of the system
(\ref{eqRde}) exist only if the factor $\alpha$ is large enough.
These limits on $\alpha$ depend on $\delta$, $\gamma$ and other
parameters.
During calculations with results, depicted in Fig.~\ref{F1}, we
excluded the mentioned oscillations by considering only solutions
with limited value of $|dR/d\log a|$. These calculations show that
the EDE term (\ref{FEDE}) with rather large $\alpha$ is the
necessary condition for viability of tis model.\\

\begin{figure}[ht]
  \centerline{\includegraphics[scale=0.68,trim=6mm 4mm 5mm 1mm]{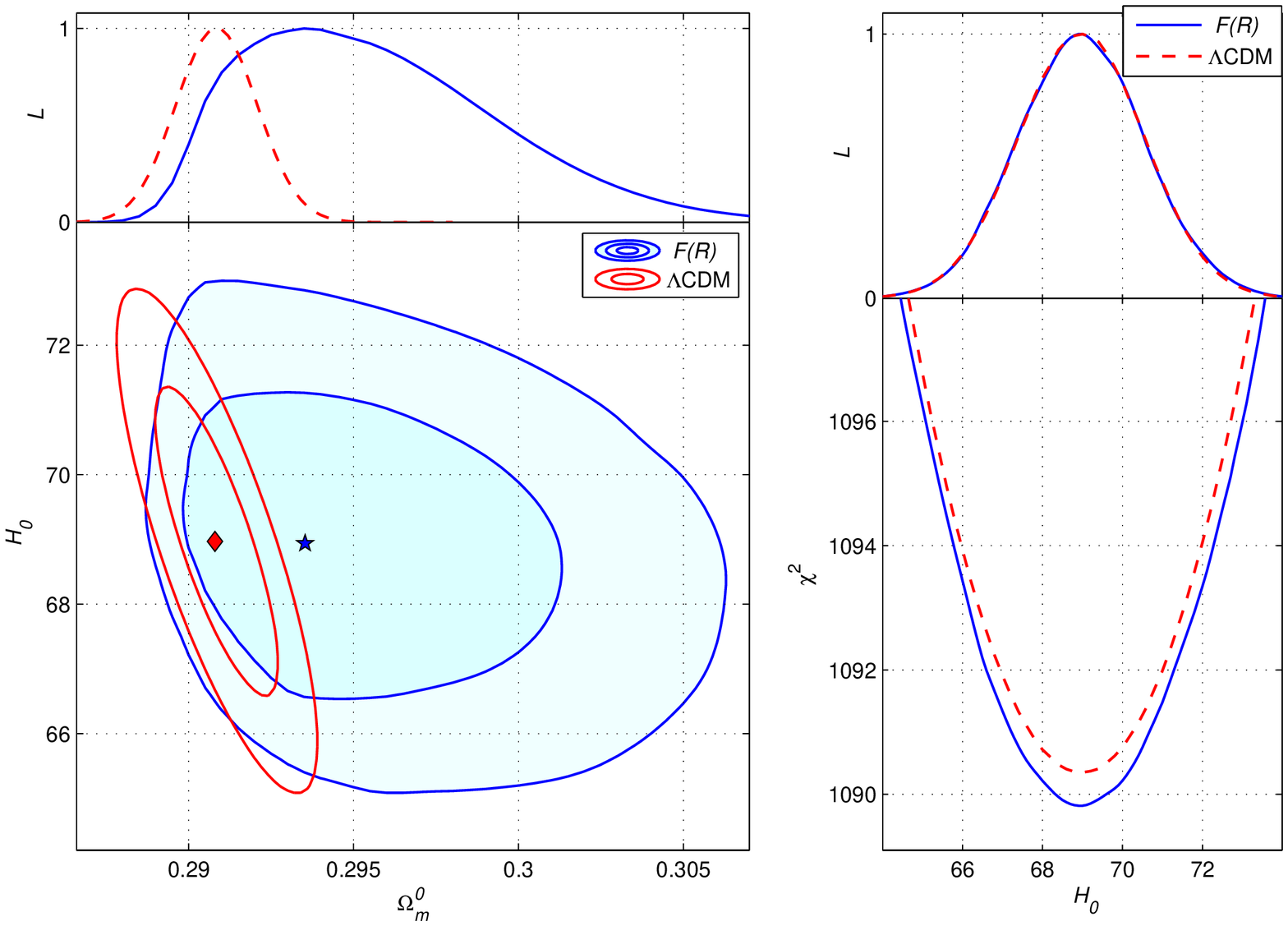}}
\caption{Contour plots of $\chi^2_\mathrm{tot}$ in $\Omega_m^0-H_0$ plane with
$1\sigma$, $2\sigma$ CL and one-parameter distributions $\chi^2_\mathrm{tot}(H_0)$
likelihood functions $ {\cal L}$ for the $F(R)$ model 
in comparison with  the $\Lambda$CDM model.}
  \label{F3}
\end{figure}

As mentioned above, regarding $R_0$, we fix the value
$R_0/(2\Lambda)=10^{8}$ corresponding to the epoch before or near
the recombination, and work with the remaining 6 parameters
(\ref{param}). Here $\Omega_\Lambda^*$ or $\Omega_\Lambda$ can be
considered as conditionally free parameters
\cite{OdintsovSGS:2021}, in particular, in the planes with
$\Omega_m^0$ we vary $\Omega_\Lambda^*$ trying to reach the
equality $E(z=0)=1$, that is equivalent $\Omega_m^*=\Omega_m^0$
and $H_0^*=H_0$. We will use this approach below, for the contour
plots in the $\Omega_m^0-H_0$ plane as shown in Fig.~\ref{F3}.
In Fig.~\ref{F3} $1\sigma$ and $2\sigma$ contour plots and
likelihood functions are presented in the $\Omega_m^0-H_0$ plane.
These parameters are common for our $F(R)$ gravity scenario and
the $\Lambda$CDM model so we can compare the corresponding
contours. The $\Lambda$CDM $1\sigma$ and $2\sigma$ CL domains are
essentially more compact and a bit shifted along the $\Omega_m^0$
axis, however the observed $H_0$ dependence is very close for both
models. It may be seen in the right panels in Fig.~\ref{F3} where
two likelihoods practically coincide and the one-parameter
distributions $\chi^2_\mathrm{tot}(H_0)$ are similar and differ
only because of the lowest minimum $\min\chi^2_\mathrm{tot}$ of
the $F(R)$ gravity model.

One can see in the figures and Table \ref{Estim}, that the
power-law $F(R)$ gravity model can describe the observational data
including SNe Ia, CC $H(z)$, CMB and BAO data with the absolute
minimum $\min\chi^2_\mathrm{tot}$ better than the $\Lambda$CDM
model. However, the $\Lambda$CDM model with $N_p=2$ free
parameters demonstrates the best AIC.

The best fit values of the Hubble constant $H_0$ and the
likelihoods ${\cal L}(H_0)$ are very close for both models (see
Fig~\ref{F3} and Table \ref{Estim}). This result confirms the
conclusion in Ref.~\cite{OdintsovSGS:2021} about weak
effectiveness of the $F(R)$ gravity model (\ref{FRde}) in
alleviating the $H_0$ tension. We see that the EDE term in the
form (\ref{FEDE}) with suitable parameters $\alpha$, $R_0$ does
not change essentially the resulting best fit of $H_0$. However
this EDE term should not be excluded from this model: it is
necessary for suppressing the above mentioned oscillations, which
inevitably appear during the epoch $10^2<z<10^3$, if  $\alpha$ is
not large enough.

The best fit values of the parameter $\delta\simeq=0.01^{+0.017}_{-0.01}$ satisfy the
condition $\delta\ll 1$ that is necessary for asymptotic behavior (\ref{asymLCDM}) of
this $\gamma R^\delta$ model. At $\delta>0.04$ this model does not describe effectively
the observational data. For $\gamma$ and the EDE factor $\alpha$ we observe rather wide
ranges of admissible values in Table \ref{Estim}.

Hence, the $F(R)$ gravity scenario (\ref{FRde}) with the EDE term
(\ref{FEDE}) gives interesting possibilities and viability in
confronting with SNe Ia, CC, BAO and CMB observational data.

\section{Conclusions}
\label{conclusions}

In this paper we have explored the power-law $F(R)$ gravity
gravity model (\ref{FRde}) with an additional EDE term
(\ref{FEDE}). Its cosmological evolution was studied by solving
dynamical equations and model predictions were confronted with
observational data including Pantheon SNe Ia, estimations of CC
$H(z)$, CMB and BAO observed manifestations.

The results are presented in Figs. \ref{F1}, \ref{F3} and  Table
\ref{Estim}. The best fit value of the Hubble constant $H_0$ for
the $F(R)$ gravity model is very close the $\Lambda$CDM model
prediction. On may conclude that the EDE term (\ref{FEDE}) does
not shift essentially the effective value of $H_0$. But we can not
exclude the EDE term $-\alpha R/({R_0+R})$ because of oscillatory
behavior of this model with $\alpha=0$ or small $\alpha=0$. These
oscillations, shown in  Fig.~\ref{F1}, can be suppressed only if
$\alpha$ is sufficiently large. If $\alpha$ is too small or
$\alpha=0$ these oscillations grow to extremely large amplitudes,
and we obtain discontinuous solutions of the system (\ref{eqRde}).
We may conclude that the EDE term (\ref{FEDE}) with  sufficiently
large $\alpha$ is necessary for viability of the model
(\ref{FRde}).

The other free parameters of the  model with the best fit values
given in Table \ref{Estim}, satisfy the conditions leading to the
$\Lambda$CDM  asymptotic behavior (\ref{asymLCDM}) before or near
the recombination epoch. In particular, for  $\delta$ the
condition $\delta\ll 1$ is fulfilled.  At $\delta>0.04$ this model
does not describe effectively the observational data. For $\gamma$
and the EDE factor $\alpha$, we observe rather wide ranges of
admissible values in Table \ref{Estim}. The estimate
$\Omega_m^0=0.294^{+0.0048}_{-0.0036}$ of the $F(R)$ model is
slightly shifted and has larger width in comparison with the
$\Lambda$CDM prediction.

We see in Figs. \ref{F1}, \ref{F3} and  Table \ref{Estim} that the
power-law $F(R)$ gravity model can successfully describe the
mentioned observations: its absolute minimum
$\min\chi^2_\mathrm{tot}$ is better in comparison with the
$\Lambda$CDM model. However, the $\Lambda$CDM model is more
optimal from the point of view of the Akaike information criterion
\cite{Akaike} and gives the smallest $\mbox{AIC} =
\min\chi^2_{tot} +2N_p$ because of the least number $N_p=2$ of its
free model parameters.

\section*{Acknowledgments}

This work was partially supported by MICINN (Spain), project
PID2019-104397GB-I00  and by the program Unidad de Excelencia
Maria de Maeztu CEX2020-001058-M, Spain (SDO).

\end{document}